\documentstyle[prl,aps,twocolumn,epsf]{revtex}

\catcode`\@=11

\def\maketitle2{\par 
\begingroup
\let\cite\@bylinecite
\def\thefootnote{\fnsymbol{footnote}}%
\twocolumn[\@maketitle2\vskip2pc]%
\thispagestyle{plain}\@thanks
\endgroup
\def\thefootnote{\arabic{footnote}}%
\setcounter{footnote}{0}%
\let\maketitle2\relax \let\@maketitle2\relax
\let\@thanks\relax \let\@authoraddress\relax \let\@title\relax
\let\@date\relax \let\thanks\relax \let\@abstract\relax 
\let\@pacs\relax}

\def\abstract#1{\gdef\@abstract{{\par 
\bgroup
\ifdim\prevdepth=-1000pt \prevdepth0pt\fi
\hsize\columnwidth
\dimen0=-\prevdepth \advance\dimen0 by17.5pt \nointerlineskip
\small\vrule width 0pt height\dimen0 \relax}{~~}#1\egroup}}

\def\pacs#1{\gdef\@pacs{{\par 
\bgroup
\hsize\columnwidth \parindent0pt
\ifdim\prevdepth=-1000pt \prevdepth0pt\fi
\dimen0=-\prevdepth \advance\dimen0 by20pt\nointerlineskip
\egroup} PACS numbers:~#1}}

\def\@maketitle2{
\@preprint
\@title
\ifdim\prevdepth=-1000pt \prevdepth0pt\fi
\@authoraddress
\@date
\begin{list}{}{\leftmargin=0.10753\textwidth \rightmargin=\leftmargin
\itemsep=1pc\partopsep=-1pc}
\item\@abstract
\item\@pacs
\end{list}
}

\catcode`\@=12

\makeatletter

\def\compoundrel#1\over#2{\mathpalette\compoundreL{{#1}\over{#2}}}
\def\compoundreL#1#2{\compoundREL#1#2}
\def\compoundREL#1#2\over#3{\mathrel
  {\vcenter{\hbox{$\m@th\buildrel{#1#2}\over{#1#3}$}}}}
\makeatother

\begin{document}

\draft

\title{Quantum limit of decoherence:
Environment induced superselection of energy eigenstates}

\author{Juan Pablo Paz$^{1}$ \thanks{paz@df.uba.ar} and
Wojciech Hubert Zurek$^{2}$\thanks{whz@t6-serv.lanl.gov}}

\address{$^1$Departamento de F\'{\i}sica J.J. Giambiagi,
FCEN, UBA, Pabell\'on 1, Ciudad Universitaria, 1428 Buenos Aires,
Argentina}

\address{$^2$Theoretical Astrophysics, MSB288, Los Alamos National
Laboratory,
Los Alamos, NM 87545, USA}


\abstract
{We investigate decoherence in the limit where the interaction with the
environment is weak and the evolution is dominated
by the self Hamiltonian of the system. We show that in this case
quantized
eigenstates of energy emerge as pointer states selected through the
predictability sieve.}

\date{\today}

%
%

\maketitle2
\narrowtext

The relevance of decoherence in the context of the quantum
to classical transition has been recently recognized
\cite{Zurek82,deco2,deco1}. The basic idea is that classicality
is an emergent property induced in open quantum systems by their
environments. Due to the interaction with the
environment, the vast majority of states in the Hilbert space of a
quantum
open system become highly unstable to entangling interaction with the
environment, which in effect monitors selected observables of the
system.
After a decoherence time, which for macroscopic objects is typically
many orders of magnitude shorter than any other dynamical timescale
\cite{Zurektime}, a generic quantum state decays
into a mixture of ``pointer states''. In this way the environment
induces effective superselection rules (``einselection'')
thus precluding stable existence of superpositions of pointer
states. Experimental testing of decoherence
has been recently initiated \cite{Bruneetal}.
Here we investigate decoherence in the limit of weak interaction and
show that it can enforce ``quantum jumps'' between discrete energy
eigenstates which now become stable pointer states.

Pointer states are distinguished by their ability to persist in spite of
the environmental monitoring
and therefore are the ones in which quantum open systems are observed.
Understanding the nature of these states and the process
of their dynamical selection is of fundamental importance.
This process has been studied first in a measurement situation:
When the system is an apparatus whose intrinsic dynamics
can be neglected, pointer states turn out to be eigenstates of the
interaction Hamiltonian between the apparatus and its environment
\cite{Zurek82}. Even in this idealized limit, decoherence may be
experimentally testable \cite{Poyatos}. In more general situations, when the
system's dynamics is relevant, einselection is more complicated. 
Pointer states result from the interplay between self--evolution 
and environmental monitoring.

To study einselection,
an operational definition of pointer states has been introduced
\cite{sieve1,ZHP}. This is the ``predictability sieve'' criterion,
based on an intuitive idea: Pointer states can be defined as the
ones which become minimally entangled with the environment in the course
of the evolution. The predictability sieve criterion is a way to quantify 
this idea by using
the following algorithmic procedure: For every initial pure state
$|\Psi\rangle$, one measures the entanglement generated dynamically
between
the system and the environment by
computing the entropy ${\cal H}_\Psi(t)=-Tr(\rho_\Psi(t)
\log\rho_\Psi(t))$
or some other measure of predictability \cite{sieve1,ZHP,sieveexamples}
from the reduced density matrix of the system $\rho_\Psi(t)$
(which is initially $\rho_\Psi(0)=|\Psi\rangle\langle\Psi|$). The
entropy is a function
of time and a functional of the initial state $|\Psi\rangle$. Pointer
states
are obtained by minimizing ${\cal H}_\Psi$ over $|\Psi\rangle$ and
demanding that the answer be robust when varying the time $t$.

The nature of pointer states has been investigated using the
predictability sieve criterion only for a limited number of examples
\cite{sieveexamples}. Apart from the already mentioned case of the
measurement situation (where pointer states are simply eigenstates of
the interaction Hamiltonian) the most notable example is that of a
quantum
Brownian particle interacting bilinearly through position with a bath
of independent harmonic oscillators. In such case
pointer states are localized in phase space,
even though the interaction Hamiltonian involves the position of the
particle \cite{ZHP}. Pointer states are the result of the interplay
between
self--evolution and interaction with the environment and turn out
to be coherent states.

One may think that the above is a generic situation. However in nature
there are systems which are not found in localized states
but in eigenstates of energy. Therefore, it is natural to
ask how can this be possible. A conceivable answer could be that
such systems (the electron in an atom, for example) are coupled to their
environments through their self--Hamiltonian \cite{Hcoupling}.
But this suggestion is easily dismissed because the form of the
interaction is rather universal and generically local \cite{Zurek82}.  

In this letter we will show that, for a large class of circumstances,
even when the
interaction between the system and the environment depends on position
(or other, essentially arbitrary observable),
the pointer states selected through decoherence turn out to
be energy eigenstates. We will illustrate this emergence of quantized
energy
as a pointer observable by using the simple but physically
relevant example of a particle interacting locally with a quantum scalar
field. The nature of pointer states --- we shall see ---
strongly depends upon the spectral
density of the environment and, when the dominant frequencies present
in the environment are slow with respect to the system's own timescale,
pointer states turn out to be eigenstates of energy.
On the other hand, only when the
environment modes include frequencies comparable or higher than the ones
associated with the system the pointer states are localized in phase
space.

We will consider the following simple model: The system is a
particle with position $\vec x$ (moving in a $N$-dimensional space) and
the environment is
a quantum scalar field $\phi$. The interaction is local and is
described by the Hamiltonian
$
H_{int}=e\phi(\vec x).
$
Expanding the scalar field in normal modes, the Hamiltonian can be
written as
$H_{int}=\int d^N{\vec k}
(h_{\vec k}\exp(i\vec k\vec x)+ h.c.)$ where the Fourier components
$h_{\vec k}$ are proportional to anihillation operators of the quantum
field (i.e., $h_{\vec k}=e\ a_{\vec k}/(2\pi)^{N/2}(2\omega_k)^{1/2}$).
More generally, we consider models in which the
particle--field interaction is slightly nonlocal taking into account
the finite extent of the particle. In this case, the interaction
Hamiltonian
$\tilde H_{int}=e\int d^N\ \vec y W(\vec x -\vec y)\phi(\vec y)$
depends upon the window function $W(\vec r)$ (whose 
support is a sphere of radius $R$, the Compton radius of the
particle,
centered around the origin). This nonlocal interaction corresponds
to a Hamiltonian whose Fourier components
$h_{\vec k}$ are multiplied by $\hat W(\vec k)$ (the transform
of $W(\vec r)$).

For this class of models we can derive a master equation for the
reduced density matrix of the particle. This equation is 
obtained under two assumptions: i) an expansion up
to second order in perturbation theory, ii) initial states with
no correlations between the system and the environment (the initial
state of the environment being thermal equilibrium).
The master equation reads \cite{masterref}:
\begin{eqnarray}
\dot\rho =-{i\over\hbar}[H,\rho]
&-&{e^2\over\hbar^2}\int d^N{\vec k}\int_0^tdt_1\nonumber\\
&\Bigl(& G_H(\vec k,t_1)\bigl[e^{i\vec k\vec x},\bigl[e^{-i\vec k\vec
x(-t_1)}
,\rho\bigr]\bigr]-\nonumber\\
&-&i G_R(\vec k,t_1)\bigl[e^{i\vec k\vec x},\bigl\{e^{-i\vec k\vec
x(-t_1)}
,\rho\bigr\}\bigr]\Bigr)\label{mother}
\end{eqnarray}
Here, $\vec x(t)$ is the Heisenberg position operator for the 
particle (evolved with the free Hamiltonian $H$) and
$G_{R,H}(\vec k,t)$ are the Fourier transform of
the retarded and symmetric two point functions 
of the scalar field (multiplied by the appropriate window function if 
the interaction is nonlocal). When the environment is a free field,
$G_R(\vec k, t)=W(\vec k)\sin(\omega_{\vec k}t)/2\omega_{\vec k}$,
$G_H(\vec k, t)=W(\vec k)\cos(\omega_{\vec k}t)(1+2N_k)/2\omega_{\vec
k}$,
where $N_k$ is the number density of particles in the 
initial state of the quantum field (the above result is valid
if the field is not free in which case the propagators are appropriately
dressed). The master equation (\ref{mother}) is extremly rich.
One of its most interesting features is that it is local in time
(note that the density matrix appearing in the r.h.s. of (\ref{mother})
is evaluated at time $t$).
We will use this equation to derive the main results of this letter.
But before, it is usefull to show how some known
results follow from Eq. (\ref{mother}).

The most widely used approximation for the particle--field
model is the so--called dipole approximation.
This is valid when the dominant wavelengths in the
environment are much larger than the lengthscale over which the position
of
the particle varies. Expanding the exponentials
up to second order ($\vec k \vec x \ll 1$) we obtain:

\begin{eqnarray}
\dot\rho=-{i\over\hbar}[H,\rho]
-{e^2\over\hbar^2}\int_0^tdt_1 &\Bigl(&
F_H(t_1)\bigl[\vec x,\bigl[\vec x(-t_1),\rho\bigr]\bigr]\nonumber\\
&-&i F_R(t_1)\bigl[\vec x,\bigl\{\vec x(-t_1),
\rho \bigr\}\bigr]\Bigr),\label{2}
\end{eqnarray}
where $F_{R,H}(t_1)=\int d^N{\vec k}
\vec k^2 G_{R,H}(\vec k,t_1)/N(2\pi)^{N/2}$.

One can recognize (\ref{2}) as the master
equation of a Brownian particle interacting (bilinearly) with an
environment
of independent oscillators. It can be further simplified
for a linear system since in this case, solving 
the free Heisenberg equations, 
$x(t)=x\ \cos(\Omega t) + {1\over m\Omega}
p \sin(\Omega t)$ for a harmonic oscillator
with frequency $\Omega$. Using this, we can rewrite the master equation
as:
\begin{eqnarray}
\dot\rho=-{i\over\hbar}\bigl[H&+&{1\over2}m\tilde\Omega^2(t)x^2,
\rho\bigr]+2i\gamma(t)\bigl[\vec x,\bigl\{\vec p,\rho\bigr\}\bigr]\nonumber\\
&-&D(t)\bigl[\vec x,\bigl[\vec x,\rho\bigr]\bigr]
 -f(t)\bigl[\vec x,\bigl[\vec p,\rho\bigr]\bigr].\label{HPZ}
\end{eqnarray}
Here the time dependent coefficients (the frequency renormalization
$\tilde\Omega(t)$, the damping coefficient $\gamma(t)$ and the two
diffusion coefficients $D(t)$ and $f(t)$) are:
\begin{eqnarray}
\tilde\Omega^2(t)&=& -{2\hbar\over m}\int_0^t dt'\cos(\Omega t')
F_R(t')\nonumber \\
\gamma(t)&=&-{\hbar\over 2m\Omega}\int_0^t dt'\sin(\Omega t') F_R(t')
\nonumber \\
D(t)&=&\int_0^t dt'\cos(\Omega t') F_H(t')\nonumber\\
f(t)&=&{1\over m\Omega}\int_0^t dt'\sin(\Omega t')
F_H(t').\label{coefHPZ}
\end{eqnarray}

The existence of a local master equation for linear Brownian motion 
was recognized some time ago \cite{HPZ}. In fact, for this system
an {\it exact} local master 
equation can always be obtained under the sole assumption of
uncorrelated initial conditions \cite{Davila}.
Equation (\ref{HPZ}) has the same form as the exact master equation
and its time dependent coefficients (\ref{coefHPZ}) coincide with the
perturbative expansion of the exact ones. The particularly simple form of
the perturbative time dependent coefficients is worth noting.
This equation has been used to study the nature of pointer states in
quantum Brownian motion models. In fact, 
if the particle moves in one dimension ($N=1$)
and the high frequency cutoff (present in $W(k)$) is much larger than
$\Omega$ we have $F_R(t)\propto \delta'(t)$. Moreover, in the limit
of high temperatures we also see that $F_H(t)\propto \delta(t)$. Using
this in equations (\ref{coefHPZ}) we find that the coefficients appearing
in the master equation are approximately constant and using this we can prove 
that pointer states tend to be localized in phase space \cite{ZHP}.  

To investigate the quantum limit of decoherence we analyze equation
(\ref{HPZ}) in the opposite regime. Thus, we
consider the case in which the frequencies
present in the environment are typically much lower than $\Omega$ (the
characteristic frequency of the system). In this case 
the kernels $F_{R,H}(t)$ varie slowly in time
and could be taken outside the integrals in (\ref{coefHPZ}). Therefore,
when the environment behaves adiabatically, the time dependent
coefficients of the master equation turn out to be 
oscillatory functions (actually, as $F_R(0)=0$ we have
$\tilde \Omega(t)\approx\gamma(t)\approx 0$, while the diffusion
coefficients oscillate
with an amplitude proportional to $F_H(0)$). Thus, one may be tempted to
conclude that for such case of an adiabatic environment the interaction
does not dynamically select any preferred basis of the system.
However, we will now show that even this very weak environment can lead to
decoherence, although it can no longer impose its own preferences for the
pointer
states: indeed, in this case they turn out to be the eigenstates of the
system Hamiltonian. To
obtain this result we take a step back and analyze the complete
master equation (\ref{mother}) without any type of dipole approximation.
The master equation (\ref{mother}) is of the form
\begin{eqnarray}
\dot\rho =-{i\over\hbar}[H,\rho]
-\sum_k &&\int_0^tdt_1 \Bigl(
 c_k(t_1)\bigl[S_k,\bigl[S_k^\dagger(-t_1), \rho\bigr]\bigr]-\nonumber
\\
&-&i c'_k(t_1)\bigl[S_k,\bigl\{S_k^\dagger(-t_1),\rho\bigr\}\bigr]\Bigr)
,\label{mother2}
\end{eqnarray}
where we have written the integral
over $\vec k$ as a sum, and introduced the operators $S_k$, which
act on the system's Hilbert space (they are
equal to $\exp(i\vec k \vec x)$) and the functions $c_k(t)$ and
$c'_k(t)$
(proportional to the Fourier transform of
the symmetric and retarded propagators of the scalar field).
When the environment behaves adiabatically we can rewrite this equation
using the following argument: On the one hand, in such case the
functions
$c_k(t)$ and $c'_k(t)$ are slowly varying and can be taken outside the
temporal integrals. On the other hand the operator $S_k(t)$ can be
written in the energy eigenbasis as:
\begin{equation}
S_k(t)=\sum_{mn} |\phi_n\rangle\langle\phi_m| S_k^{(nm)}
e^{-i\omega_{nm}t}
,\nonumber
\end{equation}
where $\omega_{nm}=(E_n-E_m)/\hbar$, $H|\phi_n\rangle=E_n|\phi_n\rangle$
and $S_k^{(nm)}=\langle\phi_n|S_k|\phi_m\rangle$.
When this expression is inserted in (\ref{mother2}) we can perform the
temporal integration and observe that all pairs of eigenstates with
different energies contribute with terms which oscillate
with frequency $\omega_{nm}$. Thus, the effect of
these terms will average out to zero. The only non--vanishing
contribution comes from terms with equal energy. Assuming that the
system has a nondegenerate spectrum the resulting master equation
(obtained from (\ref{mother2}) by averaging over the largest Bohr
period) can be written in the energy eigenbasis as 
\begin{eqnarray}
\dot\rho_{nm}=&-&i\omega_{nm} \rho_{nm}- \gamma^2_{nm} t\ \rho_{nm}
\nonumber \\
&-&t  \sum_{l\neq n,m} \bigl(A_{lnm} \rho_{lm}+B_{lnm}\rho_{bn}\bigr),
\label{mother3}
\end{eqnarray}
where $\gamma^2_{nm}$, 
$A_{lnm}$ and 
$B_{lnm}$ depend on $S_k^{(lm)}$, $c_k$ and $c'_k$.

This equation enables us to derive our main result.
On the one hand we can see that an environment behaving adiabatically
does not produce any change in the population of energy eigenstates.
Thus,
according to equation (\ref{mother3}) diagonal elements of the density
matrix do not change at all. On the other hand the evolution of
nondiagonal elements is dominated by the second term on the right hand
side of (\ref{mother2}) (notice that this term has a definite sign)
which
implies that they decay at a rate $\gamma_{nm}$ which is determined by
the
sum of the squared differences between the corresponding expectation
values
of the operators $S_k$. Thus, neglecting the contribution of the last
terms
in (\ref{mother3}) (they
will generally have alternating signs and therefore a negligible net
effect), we find that
$\rho_{nm}(t)\approx \rho_{nm}(0)\exp(-i\omega_{nm}t)
\exp(-t^2 \gamma^2_{nm})$. It is now straightforward to see that
energy eigenstates are perfect pointer states and (at the
approximation level adopted above) {\it produce no entropy}.
Thus, we have shown that energy eigenstates are the ones which are
selected by the environment as the pointer states when the environment
behaves adiabatically. It is worth pointing out that, as the decay rate
for nondiagonal elements depends on the difference between the
corresponding diagonal elements of the operators $S_k$ in the energy
eigenbasis, these operators must satisfy a simple condition for this
mechanism to take place: they must have nonvanishing diagonal elements
in the energy eigenbasis. This is the reason why this effect could
not be found using the usual master equation for quantum Brownian
motion,
which is obtained by applying the dipole approximation to equation
(\ref{mother}): In fact, in that case the operators $S_k$ are
approximated
as $S_k\approx 1-i\vec k \vec x$, whose $\vec x$--dependent part has
vanishing diagonal elements in energy eigenbasis (for the case
of the linear Brownian model discussed above).

There are three basically distinct regimes in which one can analyze the
properties of the pointer states selected through decoherence.
They differ through the relative strength of the self--Hamiltonian and
of the Hamiltonian of interaction. The
first one is the measurement situation \cite{Zurek82} where the
self--Hamiltonian of the system is negligible and the
evolution is completely dominated by the interaction with the
environment. In such case, pointer states are eigenstates
of the interaction Hamiltonian. The second, most common and complex
situation
occurs when neither the self--Hamiltonian nor the interaction with
the environment are clearly dominant. Then the pointer states arise
from a compromise between self--evolution and interaction. The
most widely studied example of this situation is the QBM model for
which pointer states become localized in phase space. The third
situation completing this picture is the one we analyzed in this letter.
It corresponds to the case where the dynamics is dominated by the
system's self--Hamiltonian. In this case einselection produces
pointer states which coincide with the energy eigenstates of the
self--Hamiltonian.

Our conclusion conforms with the
heuristic picture of decoherence and einselection: The environment
``monitors'' certain states and, by doing so, elevates them to the
pointer state status. In absence of the self--Hamiltonian this leads
to selection of eigenstates of the interaction Hamiltonian. However when
energy eigenstates are separated by more than the highest energies
present in the environment a ``protective monitoring'' of energy
eigenstates will ensue \cite{Aharonov,Unruh}. Thus, an environment
coupling to the system through nearly any observable will 
become correlated with an energy eigenstate because time--average of any
observable over a time necessary to establish correlations (achieving
orthogonality of records imprinted in the environment) can
depend on the only non--oscillating quantity: energy.

Quantum jumps are a proverbial characteristic of quantum theory, and an
old subject of heated debates. Schr\"odinger had hoped that his equation
will do away with the discreteness, replacing jumps with a comprehensible
or at least continuous process. He became physically ill when Bohr convinced
him otherwise, and was never reconciled with this conclusion \cite{Sch52}.
The jumps were a phenomenological, but phenomenally succesfull 
rule-of-thumb introduced by Bohr in the old quantum theory. 
However, in absence of some sort of collapse postulate they are difficult to
understand within the purely unitary evolution of a closed quantum
system, as it is illustrated by recent exchanges of comments \cite{Zehjumps}. 
We have now seen in one --- quite generic --- set of circumstances how 
the tell-tale quantum discreteness emerges when the continuum of 
Schr\"odinger evolution is sieved out by einselection. Remarkably, even the 
discreteness of quantum physics appears to be in part traceable to decoherence.

We have shown how Schr\"odinger equation alone suffices to bring about
discreteness by enforcing einslection of energy eigenstates, providing that
the environment is included in the considerations. Granted, the 
discrete spectrum of the self-Hamiltonian is a necessary, but --- alone ---
an insufficient condition: Superposition principle would demand "equal rights''
for arbitrary superpositions of energy eigenstates. Yet, in the presence of
an adiabatic environment, eigenstates of the self-Hamiltonian are selected 
assuming the role of pointer states.

We  thank the Benasque Center for Physics where this
work was completed. JPP was supported by Fundaci\'on Antorchas,
UBACyT, Conicet and ANPCyT. WHZ was supported by DOE grant W-7405-ENG-36.

%
%

\end{document}